\documentclass[aps,prl,amsmath,amssymb,superscriptaddress,floatfix,showpacs,twocolumn]{revtex4-1}

\usepackage{tikz}
\usepackage{verbatim}
\usepackage{bm}
\usepackage{dcolumn}
\usepackage{amsmath}
\usepackage{amsfonts}
\usepackage{amssymb}

\usepackage{graphicx}
\usepackage{epstopdf}
\usepackage{subfigure}

\usepackage{wasysym}

\usepackage{booktabs}

\usepackage{color}

\definecolor{darkblue}{rgb}{0,0.02,0.45}
\definecolor{darkred}{rgb}{0.45,0.02,0} 

\usepackage[colorlinks,bookmarks=false,citecolor=blue,linkcolor=red,urlcolor=blue]{hyperref}

\newcommand{\be}{\begin{equation}}
\newcommand{\ee}{\end{equation}}
\newcommand{\bea}{\begin{eqnarray}}
\newcommand{\eea}{\end{eqnarray}}
\newcolumntype{.}{D{.}{.}{-1}}

\def\bs{\boldsymbol}
\def\vec{\mathbf}
\def\mc{\mathcal}

\allowdisplaybreaks

\newcommand{\cuseo}{Cu$_2$OSeO$_3$}

\begin{document}

\title{Entangled tetrahedron ground state and excitations of the magneto-electric skyrmion material Cu$_2$OSeO$_3$}

\author{Judit Romh\'anyi}
\affiliation{Institute for Theoretical Solid State Physics, IFW Dresden, D-01069, Germany}
\author{Jeroen van den Brink}
\affiliation{Institute for Theoretical Solid State Physics, IFW Dresden, D-01069, Germany}
\affiliation{Department of Physics, TU Dresden, D-01062 Dresden, Germany}
\author{Ioannis Rousochatzakis}
\affiliation{Institute for Theoretical Solid State Physics, IFW Dresden, D-01069, Germany}

\date{\today}

\begin{abstract}   
The strongly correlated cuprate \cuseo\ has recently been identified as the first insulating system exhibiting a skyrmion lattice phase. Using a microscopic multi-boson theory for its magnetic ground state and excitations, we establish the presence of two distinct types of modes: a low energy manifold that includes a gapless Goldstone mode and a set of weakly dispersive high-energy magnons. These spectral features are the most direct signatures of the fact that the essential magnetic building blocks of \cuseo\ are {\it not} individual Cu spins, but rather weakly-coupled Cu$_4$ tetrahedra. Several of the calculated excitation energies are in excellent agreement with terahertz electron spin resonance, Raman, and far-infrared experiments, while the magneto-electric effect determined within the present quantum-mechanical framework is also fully consistent with experiments, giving strong evidence in the entangled Cu$_4$ tetrahedra picture of \cuseo. The predicted energy and momentum dependence of the dipole and quadrupole spin structure factors call for further experimental tests of this picture. 
\end{abstract}

\pacs{
75.10.-b  %General theory and models of magnetic ordering
75.85.+t %Magnetoelectric effects, multiferroics (for multiferroics and magnetoelectric films, see 77.55.Nv)
76.50.+g %Ferromagnetic, antiferromagnetic, and ferrimagnetic resonances; spin-wave resonance (see also 75.30.Ds Spin waves)
}

\maketitle

{\it Introduction---} 
The experimental discovery of skyrmions in the chiral metallic helimagnets MnSi~\cite{Muhlbauer2009,tonomura2012}, Fe$_{1-x}$Co$_x$Si~\cite{Yu2010,Muenzer2010} and FeGe~\cite{yu2012}, almost 20 years after their theoretical prediction~\cite{Bogdanov1989a,Bogdanov1989b}, has prompted an enormous interest in the community, both on the theory and the experiment side. Skyrmions are localized magnetization textures with non-trivial topology which, under certain conditions~\cite{Bogdanov1989a,Bogdanov1989b,Roessler2006}, may condense into a lattice, in analogy to the Abrikosov vortices in type-II superconductors~\cite{Abrikosov1957}, and the so-called ``blue phases'' in chiral liquid crystals~\cite{Wright89}. The recent discovery of skyrmions~\cite{Seki2012Sc} in \cuseo\  has opened an alternative route for realizing this kind of physics, since this material is an insulating oxide with localized spins-$1/2$~\cite{Bos2008}. Moreover \cuseo\ shows a linear magneto-electric (ME) effect~\cite{Belesi2012,Seki2012Sc,SekiarXiv} which in principle allows the skyrmions to be manipulated by an external electric field without losses due to joule heating~\cite{White2012,Lin2013,Fert2013}. 

\begin{figure}
\begin{center}
\includegraphics[width=\columnwidth]{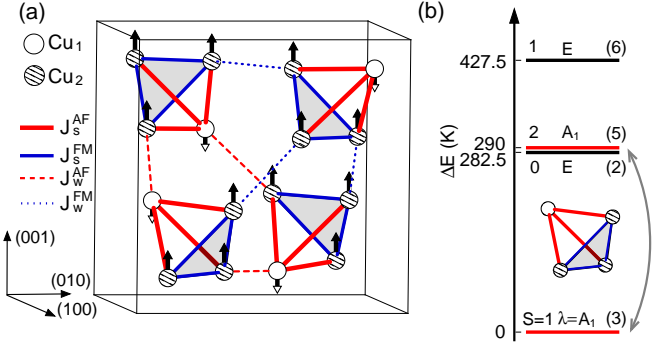}
\caption{(color online) (a) Distorted pyrochlore structure of \cuseo. Solid (dashed) lines indicate the strong (weak) exchange couplings. For clarity, the longer-range $J^{AF}_{\sf O..O}$ coupling (see text) is not shown here. (b) The energy spectrum of an isolated strong tetrahedron~\cite{suppl}.}
\label{fig:structure}
\end{center}
\end{figure}

A microscopic understanding of this rich set of physical phenomena in Mott insulating \cuseo\  starts from a  description of its magnetic interactions. Close inspection of  the superexchange interactions between localized Cu$^{2+}$ spins, as calculated from atomistic {\it ab initio} methods~\cite{Yang2012,Janson2013}, reveals the presence of weaker and stronger magnetic bonds. Due to this separation of energy scales, the degrees of freedom that order below T$_\text{C}\simeq 60$~K and twist in the helical or the skyrmion lattice phase are {\it not} the bare Cu$^{2+}$ spins but instead tetrahedral entities (see Fig.~\ref{fig:structure}(b)), which are, nevertheless, also deeply quantum-mechanical (QM) in nature~\cite{Janson2013}.
These entities constitute the basic building blocks of the helimagnetism and the magnetoelectricity (see below) of \cuseo\ while, at the same time, they affect very strongly a number of key quantities of direct experimental interest, such as: the local spin lengths, the ordering temperature $T_C$, the exchange stiffness, the twisting parameter, the diameter of the skyrmions, the presence of a weak antiferromagnetic modulation of the primary order parameter, and the sign of the magnetic handedness~\cite{Janson2013}.

Here we demonstrate that the notion of weakly-interacting Cu$_4$ tetrahedra gloriously survives a full quantitative QM treatment of the problem and has profound implications for the magnetic excitation spectrum as well as for the ME properties of \cuseo. Using a multi-boson approach that incorporates the strong tetrahedral entities  from the outset --in this way capturing the dominant portion of the QM correlations-- we establish the presence of two distinct types of excitations: (i) a 4-branch low-energy manifold that includes a gapless Goldstone mode corresponding to long-wavelength modulations of the local order parameter, and (ii) a higher-energy set of more weakly dispersive intra-tetrahedron excitations. 
A number of excitation energies at zero momentum are in striking agreement with reported terahertz electron spin resonance (ESR)~\cite{ozerov2014}, Raman~\cite{Gnezdilov2010}, and far-infrared~\cite{Miller2010} data, lending strong support to the entangled tetrahedra picture of this compound. Given that this picture can be also tested by other spectroscopic techniques, such as inelastic neutron scattering, we explicitly calculate the dipolar and quadrupolar dynamical structure factors for the entire spectrum. We also explore the ramifications of this framework for the ME coupling by calculating the induced electric polarization $\vec{P}$ in different crystallographic directions as a function of the orientation of an applied magnetic field $\vec{H}$. Recent experimental observations~\cite{Belesi2012, Seki2012Sc, SekiarXiv} are fully consistent with the calculated direction of $\vec{P}$, again supporting the fundamental notion that the essential magnetic building blocks of \cuseo\  are not individual Cu spins, but entangled Cu$_4$ tetrahedra. 

{\it Magnetic ground state---}
\cuseo\ has the non-centrosymmetric P2$_1$3 space group, similar to the metallic B20 helimagnets. There are two symmetry inequivalent Cu$^{2+}$ sites, Cu$_1$ and Cu$_2$, residing at the Wyckoff positions 4a and 12b, respectively~\cite{Bos2008,Belesi2010,Gnezdilov2010}. These sites form a distorted, 3D pyrochlore lattice of corner sharing tetrahedra, consisting of one Cu$_1$ and three Cu$_2$ ions each, see Fig.~\ref{fig:structure}(a). For present purposes, we consider only the Heisenberg-type exchange interactions described by the Hamiltonian $\mc{H}_{\text{Heis}}=\sum_{\langle ij\rangle}J_{ij}{\bf S}_i\cdot{\bf S}_j$ (where $J_{ij}$ are the exchange couplings between sites $i$ and $j$). 
The weaker Dzyaloshinskii-Moriya interactions~\cite{Yang2012,Janson2013} affect only the very low frequency portion of the excitation spectrum close to the $\Gamma$ point, as shown by the agreement to experiment below and by the very low value of the twisting parameter of the effective action~\cite{Janson2013}.
%The much weaker Dzyaloshinskii-Moriya (DM) interactions~\cite{Yang2012,Janson2013} affect only the very low frequency portion of the excitation spectrum close to the $\Gamma$ point. 
%
%
Four of the five most relevant exchange paths are indicated in Fig.~\ref{fig:structure}: $J^{\sf AF}_{\sf s}\!=\!145$K, $J^{\sf FM}_{\sf s}\!=\!-140$K (within strong tetrahedra, solid lines) and $J^{\sf AF}_{\sf w}\!=\!27$K and $J^{\sf FM}_{\sf w}\!=\!-50$K (connecting strong tetrahedra, dashed lines). The fifth coupling is a longer-range exchange, $J^{\sf AF}_{\sf O..O}\!=\!45$K, connecting Cu$_1$ with Cu$_2$ sites across the diagonals of alternating Cu$_1$--Cu$_2$ hexagon loops. 
The weak coupling values are the ones obtained in [\citenum{Janson2013}] based on {\it ab initio} calculations and subsequent comparison to experimental magnetization data, while the values of $J^{\sf AF}_{\sf s}$ and $J^{\sf FM}_{\sf s}$ are the ones extracted from the recent terahertz ESR data~\cite{ozerov2014} using the present theoretical framework, and are different from the ones given in [\citenum{Janson2013}] by only 10-15\,\%.

The presence of stronger ($J^{\sf AF}_{\sf s}$ and $J^{\sf FM}_{\sf s}$) and weaker couplings ($J^{\sf AF}_{\sf w}$, $J^{\sf FM}_{\sf w}$, and $J^{\sf AF}_{\sf O..O}$)  suggests that we take as a starting point a tetrahedron--factorized wave function $|\Psi\rangle=\prod_{t}^\otimes |\psi\rangle_{t}$, where $|\psi\rangle_t$ is a QM state living in the 16-dimensional Hilbert space of the strong tetrahedron $t$. Doing so is equivalent to solving the single tetrahedron mean field (TMF) Hamiltonian~\cite{Janson2013} $\mc{H}_{\text{TMF}}^{(t)}\!=\!\mc{H}_0^{(t)}\!+\!\mc{V}_{\text{MF}}^{(t)}$, where $\mc{H}_0^{(t)}$ contains the intra-tetrahedra couplings, and $\mc{V}_{\text{MF}}^{(t)}$ the exchange fields exerted from $t'\neq t$ at a mean field level. The ensuing energy eigenstates of $\mc{H}^{(t)}_0$ are shown in Fig.~\ref{fig:structure}(b) above, and can be labeled by the total spin $S$, its projection $S_z$, and the irreducible representations $\lambda$ of the point group C$_{3\text{v}}$. The ground state is an A$_1$--triplet with a large excitation gap of $\sim 282.5$~K. A finite $\mc{V}_{\text{MF}}$ mixes states with different $S$, so that it is no longer a good quantum number. The point group symmetry remains C$_{3\text{v}}$, however, thus the A$_1$--triplet can only admix with the A$_1$--quintet. So for an infinitesimal symmetry breaking magnetic field along $\vec{z}$, the ground state of $\mc{H}_{\text{TMF}}^{(t)}$ reads
\begin{eqnarray}
|\psi\rangle_t \!=\! \cos\frac{\alpha}{2}~|1,1,A_1\rangle_t + \sin\frac{\alpha}{2}~|2,1,A_1\rangle_t~,
\label{eq:vari_coherent}
\end{eqnarray}
where the variational parameter $\alpha$ controls the degree of spin mixing and the local moments, since  
\be\label{eq:S1S2} 
\langle S_1^z\rangle\!=\!-\frac{1}{4}(\cos\alpha+\sqrt{3}\sin\alpha), ~~ \langle S_2^z\rangle\!=\!\frac{1}{3}(1 -\langle S^z_1\rangle)~.
\ee
Incidentally, the total moment per tetrahedron is $\langle S_1^z\rangle\!+\!3\langle S_2^z\rangle\!=\!1$ regardless of $\alpha$, which corresponds to a 1/2 magnetization plateau. For $\alpha\!=\!0$, $\langle S_1^z\rangle\!=\!-\frac{1}{4}$ and $\langle S_{2}^z\rangle\!=\!\frac{5}{12}$, while in the coupled limit the minimization of $_t\langle\psi| \mc{H}_{\text{TMF}}^{(t)} |\psi\rangle_t$ yields $\alpha\!=\!0.337205$, $\langle S^z_1\rangle\!\simeq\!-0.38$ and $\langle S^z_{2}\rangle\!\simeq\!0.46$. The reduced values of the spin lengths compared to the classical $S_{1,2}^z\!=\!\mp\frac12$ values reflect the fact that $|\psi\rangle$ is highly entangled and cannot be decomposed into a site-factorized form. Such a local moment reduction has indeed been observed experimentally~\cite{Bos2008}.

\begin{figure*}
\begin{center}
\includegraphics[width=1.99\columnwidth]{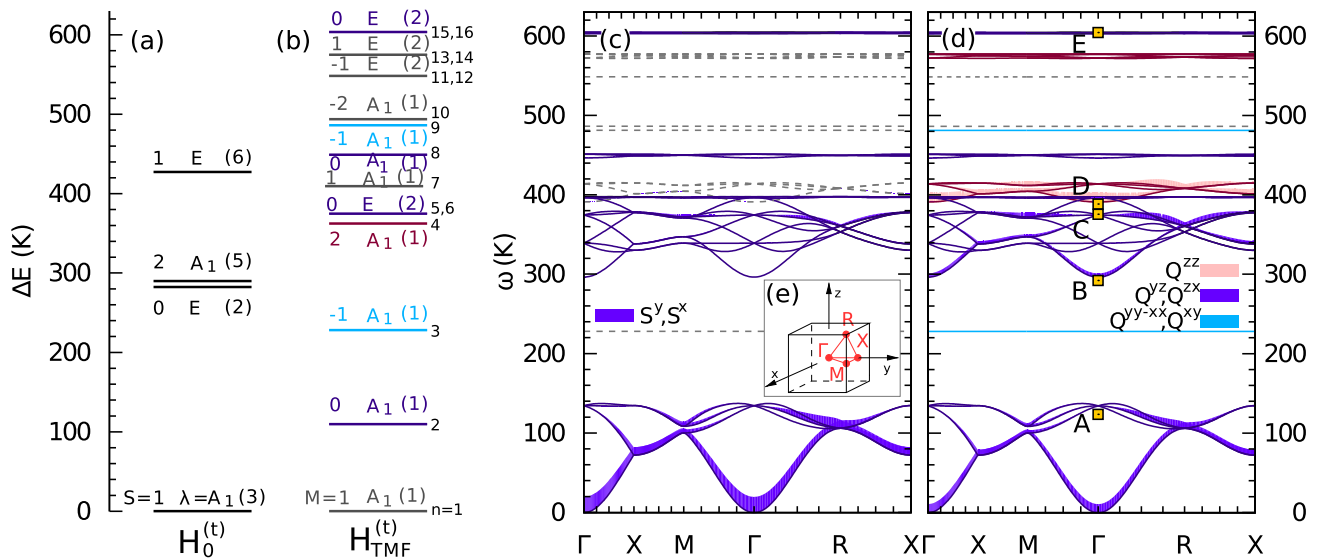}
\caption{(color online) 
(a-b) Energy levels of $\mc{H}_0^{(t)}$ and $\mc{H}^{(t)}_{\sf TMF}$, see text. 
(c,d) Multi-boson spectrum of Cu$_2$OSeO$_3$. The width of each line represents the matrix elements of the magnetic dipole (c) and quadrupole (d) operators. 
Dashed lines correspond to vanishing matrix elements due to selection rules. Yellow squares A-E in (d) indicate the five Raman lines reported in [\onlinecite{Gnezdilov2010}], see text. 
(e) Brillouin-zone path taken in (c) and (d).
}
\label{fig:spectrum}
\end{center}
\end{figure*}

{\it Magnetic fluctuations and excitations---}
Owing to the QM nature of $|\Psi\rangle$, including the fluctuating part $\mc{H}\!-\!\sum_t\mc{H}_{\text{TMF}}^{(t)}$ requires a multi-boson generalization~\cite{Papanicolaou1984,Papanicolaou1988,Onufrieva1985,Chubukov1990} of the standard spin-wave expansion~\cite{Holstein1940,Anderson1952,Kubo1952}. Such multi-boson theories have previously been employed successfully in quantum spin models with dimerized~\cite{Romhanyi2011}, quadrupolar~\cite{Carretta2010,Shiina2003}, or nematic phases~\cite{nematicreview2011,Blume1969}.
 
We first introduce bosonic operators $a_{n,\nu}(\vec{R})$, with $n\!=\!1$-$16$ and $\nu\!=\!1$-$4$, such that $a^{\dagger}_{n,\nu}(\vec{R})|0\rangle$, where $|0\rangle$ is the vacuum, gives the $n$-th eigenstate of $\mc{H}_{\text{TMF}}$ [see Fig.~\ref{fig:spectrum}(b)] at the $\nu$-th tetrahedron inside the unit cell $\vec{R}$. The wavefunction $|\Psi\rangle$ can be thought of as a coherent state $\prod_{\nu,\vec{R}} (a^{\dagger}_{1,\nu}(\vec{R}))^M|0\rangle$ as $M\!\to\!\infty$. The remaining 15 bosons per tetrahedron play the role of generalized ``tetrahedral spin waves''. Now, given that each tetrahedron can only be in one of the 16 states $|n\rangle$, the bosons must satisfy the hard-core constraint $\sum_{n} a^{\dagger}_{n,\nu}(\vec{R}) a_{n,\nu}(\vec{R})\!=\!1$. Following the standard approach~\cite{Holstein1940}, to treat this constraint we allow for $M$ bosons per tetrahedron instead of $1$, and replace the condensed bosons with 
\bea
a_{1,\nu} \!=\! a^\dagger_{1,\nu} \!=\! \sqrt{M-\sum_{n\!>\!1} a^\dagger_{n,\nu} a_{n,\nu}}~, \label{eq:repl}
\eea
and then perform a $1/M$ expansion up to quadratic terms. In the bosonic Hamiltonian, the zeroth-order scalar corresponds to the mean-field energy, the first-order terms vanish in the ground state, while the second-order terms provide the quadratic spectrum by a standard Bogoliubov diagonalization~\cite{suppl}.

This calculation provides direct access to the residual spin length corrections, which are an important diagnostic for the reliability of the multi-boson approach. It quantifies the strength of residual quantum fluctuations, which should be small for our theoretical approach to be consistent. We find (see SM for details) $\delta S_1\!=\!0.0137423$ and $\delta S_2\!=\!-0.00458264$, which are very small indeed. This implies that most of the quantum correlations have already been captured by $|\Psi\rangle$ and that  tetrahedral spin-wave fluctuations have a negligible influence on the spin lengths: our multi-boson expansion is thus based on very firm ground.

The resulting magnetic excitation spectrum is shown in Figs.~\ref{fig:spectrum}(c-d). Most of the features can be understood by comparing with the energy levels of $\mc{H}_{\text{TMF}}$ for $\alpha\!=\!0.337205$, show in Fig.~\ref{fig:spectrum}(b). The lowest excitations of $\mc{H}_{\text{TMF}}$ from $n\!=\!1$ to $n\!=\!2$ give rise to four modes (counting the number of tetrahedra per unit cell) with $\delta M\!=\!1$. One of them gives the quadratic Goldstone mode (associated with the breaking of SU(2) symmetry down to U(1)), and the remaining three modes are at $\sim 134.5$~K at the $\Gamma$ point. The excitation  from $n\!=\!1$ to $n\!=\!3$ gives another four modes with $\delta M\!=\!2$, which sit at $\sim 228$~K and are dispersionless. Excitations from $n\!=\!1$ to $n\!\ge\!4$ give rise to higher energy modes above $\sim 296.5$~K, which are all weakly dispersive.

{\it Comparison to experiments ---} 
Gnezdillov {\it et al.}~\cite{Gnezdilov2010} have reported five possible magnetic modes in the energy scale of interest from Raman experiments. 
These modes are shown by square symbols on top of the calculated spectrum in Fig.~\ref{fig:spectrum}(c), and are labeled by A (86 cm$^{-1}$), B (203 cm$^{-1}$),  C (261 cm$^{-1}$), D (270 cm$^{-1}$), and E (420 cm$^{-1}$). The lines A, B and D have also been observed in far-infrared absorption spectra by Miller {\it et al}~\cite{Miller2010}, but the magnetic nature of line A was debated. 

Our theoretical picture and the magnetic nature of most of the above modes has been firmly established recently by the ESR experiments by Ozerov {\it et al.}~\cite{ozerov2014}. Using a terahertz free electron laser and high magnetic fields up to 60~T, these experiments have given direct access not only to the expected long-wavelength Goldstone mode, but also to the finite energy modes A~\cite{modeA}, B and C. The consistency between three different experimental studies and the quantitative agreement with our calculations give very strong confidence to the entangled tetrahedra picture of \cuseo.

{\it Dynamical structure factors ---} 
To give predictions for further experiments, we calculate the dipole and quadrupole matrix elements between the magnetic ground state $|i\rangle$ and excited states $|f\rangle$. In terms of the magnetic dipole operators, the quadrupolar operators, defined on bonds $(ij)$, are $Q^{\beta\gamma}_{ij}\equiv S^\beta_{i} S^\gamma_{j} + S^\gamma_{i} S^\beta_{j}\equiv\overline{S^\beta_{i} S^\gamma_{j}}$, where $\beta,\gamma$ are cartesian coordinates. For simplicity we restrict ourselves here to the ones on the strong bonds: 
\bea
&& Q^{\beta\gamma}_{{\sf s, AF}}(\vec{R})\!=\!\overline{S^\beta_{1,t} S^\gamma_{2,t}} + \overline{S^\beta_{1,t} S^\gamma_{3,t}} + \overline{S^\beta_{1,t} S^\gamma_{4,t}}\,, \nonumber\\
&&Q^{\beta\gamma}_{{\sf s,FM}}(\vec{R})\!=\!\overline{S^\beta_{2,t} S^\gamma_{3,t}} + \overline{S^\beta_{3,t} S^\gamma_{4,t}} + \overline{S^\beta_{4,t} S^\gamma_{2,t}}\,, \label{eq:Qs}
\eea
where $t$ is any of the four tetrahedra in the unit cell $\vec{R}$.
The dipole matrix elements are directly relevant for inelastic neutron scattering (INS) experiments, since the INS intensity is proportional to the dynamical spin structure factor $S^{\mu\nu}({\bf k},\omega)=\sum_f\langle i|S^\mu_{-{\bf k}}|f\rangle\langle f|S^\nu_{\bf k}|i\rangle\delta(\omega-\omega_{fi})$. In addition, these results are relevant in light scattering experiments, since the response to the electric and magnetic components of the incoming light is given by  the magnetic and dielectric susceptibilities, $\Im\chi^{mm}_{\alpha\alpha}(\omega)\propto\sum_f |\langle i|S^\alpha|f\rangle|^2\delta(\omega-\omega_{fi})$ and $\Im\chi^{ee}_{\alpha\alpha}(\omega)\propto\sum_f |\langle i|P^\alpha|f\rangle|^2\delta(\omega-\omega_{fi})$, where $\vec{P}$ is the electric polarization. The latter is related to quadrupolar spin operators though the ME coupling~\cite{Belesi2012} (see also below). 

In Fig.~\ref{fig:spectrum}, the magnetic dipole transition matrix elements (c) and the quadrupole ones $Q^{\beta\gamma}\!=\!Q^{\beta\gamma}_{{\sf s,AF}}\!+\!Q^{\beta\gamma}_{{\sf s,FM}}$ (d) are indicated by the width of the lines in the spectrum. Dashed lines denote matrix elements that vanish exactly due to selection rules, as e.g. in the dipole case for transitions with $\delta M\!=\!2$ (such transitions are instead allowed in $Q^{xx-yy}$ and $Q^{xy}$). 
The dipole transitions are particularly strong for the lowest four branches, whereas the quadrupolar ones have appreciable weights at both low and high energy modes. 

The fundamental difference between inter- and intra-tetrahedra excitations is directly reflected in the characteristic dispersion in momentum space and is thus a key element for experimental comparisons: The dispersion is more pronounced for the low-energy modes and becomes progressively weaker as we reach the high-energy end of the spectrum.
In addition, establishing experimentally the value of the exchange stiffness for the long-wavelength Goldstone mode would provide one of the two crucial parameters of the long-wavelength helimagnetism (the other being the DM twisting parameter)~\cite{Janson2013}.

\begin{figure}
\begin{center}
\includegraphics[width=8cm]{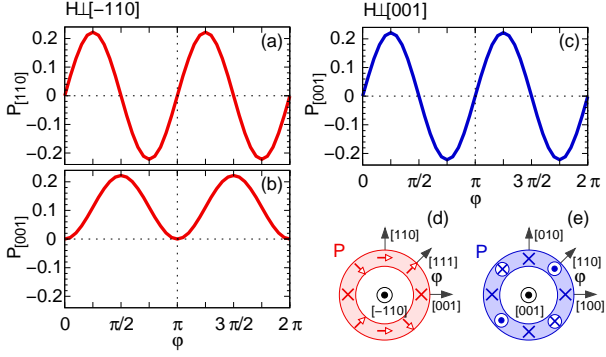}
\caption{(color online) (a) The $[110]$ and (b) the $[001]$ components of the induced electric polarization in magnetic field rotating  perpendicular to the $[\overline{1}10]$ axis. (c) $P_{[001]}$ in rotating field about the  $[001]$ axis. (d)-(e) The corresponding geometries, chosen according to the convention of Ref.~[\onlinecite{SekiarXiv}].}
\label{fig:induced_P}
\end{center}
\end{figure}

{\it Magneto-electric effect} ---
As the quadratic corrections to the state $|\Psi\rangle$ are extremely small, $|\Psi\rangle$ provides accurate information about other GS expectation values, such as for instance those of quadrupole spin operators. The coarse-grained electric polarization $\vec{P}$~\cite{Belesi2012}, in particular, can be expressed as a sum over quadrupolar spin contributions from symmetry inequivalent bonds: 
\be
P^\alpha(\vec{R})/\chi_e \!=\!  \xi_{{\sf s,AF}} \langle Q^{\beta\gamma}_{{\sf s,AF}}(\vec{R})\rangle \!+\! \xi_{{\sf s,FM}} \langle Q^{\beta\gamma}_{{\sf s,FM}}(\vec{R})\rangle \!+\! \ldots \,,~~~
\ee
where $\beta\gamma\!=\!yz$, $zx$, $xy$ for $\alpha\!=\!x$, $y$, $z$, respectively, $\chi_e$ is the dielectric susceptibility, and $\xi_{{\sf s,AF}}$, $\xi_{{\sf s,FM}}$ are ME coupling constants. 
We emphasize here that the phase space of isolated spins-1/2 leaves no room for quadrupole moments, and thus the minimal ME coupling must involve at least two Cu$^{2+}$ spins, excluding single-ion mechanisms. The symmetric combinations of (\ref{eq:Qs}) are picked out by the symmetry of $|\Psi\rangle$, thus reflecting the special role of the entangled Cu$_4$ entities for the ME effect. 

For a magnetic field pointing along $\bs{\Omega}\!=\!(\sin\vartheta \cos\varphi$, $\sin\vartheta \sin\varphi$, $\cos\vartheta)$, we find 
\bea
\vec{P}/\chi_e \!=\! P \left( \sin\varphi\sin2\vartheta, \cos\varphi\sin2\vartheta, \sin2\varphi\sin^2\vartheta \right) \,,
\eea
where $P\!=\!\xi_{{\sf s,AF}}Q_{{\sf s,AF}}+\xi_{{\sf s,FM}}Q_{{\sf s,FM}}+\ldots$, and
\bea
&& Q_{{\sf s,AF}}\!=\!\frac{1}{8}(-2+\cos\alpha-3\sqrt{3}\sin\alpha) = -0.346\,,\\
&& Q_{{\sf s,FM}}\!=\!\frac{3}{8}(\cos\alpha+\sqrt{3}\sin\alpha) =  0.568\,,\\
&& Q_{{\sf w,AF}}\!=\!Q_{{\sf O...O,AF}}\!=\! \langle S_1^z\rangle\langle S_2^z\rangle=  -0.173\,,\\
&& Q_{{\sf w,FM}}\!=\! \langle S_2^z\rangle^2 = 0.210\,,
\eea
where $\langle S_{1,2}^z\rangle$ were given in Eq.~(\ref{eq:S1S2}) above. In Fig.~\ref{fig:induced_P} the resulting electric polarization $\vec{P}$ along different crystallographic directions is plotted as a function of the orientation of applied magnetic field $\vec{H}$. The calculated direction of $\vec{P}$ is fully consistent with experiments~\cite{Seki2012Sc,Belesi2012,SekiarXiv}.

{\it Conclusions ---}
From a qualitative, empirical standpoint it is well-known that a broken inversion symmetry can empower weak Dzyaloshinskii-Moriya interactions --in presence of an applied magnetic field and at temperatures close to the ferromagnetic T$_{\rm C}$-- to induce a long-range ordered skyrmion lattice phase. From the theory presented here we conclude that in \cuseo\ the magnetic degrees of freedom that order and twist into a skyrmionic lattice are rigid tetrahedral entities that are highly entangled and quantum mechanical in nature. The calculations show that the tetrahedral entities are protected by a substantial gap and give rise to a rich magnetic excitation spectrum with a distinct separation of energy scales, comprising a low-energy manifold and a set of weakly dispersive high-energy magnons. This structure bears the fingerprints of the tetrahedral entities, and so the striking agreement with three experimental studies (terahertz ESR, Raman, and far-infrared) gives strong support of the fundamental notion that the essential magnetic building blocks of \cuseo\  are not individual copper spins, but entangled Cu$_4$ tetrahedral entities. The dispersions associated to this picture can be further probed by spectroscopic techniques (such as inelastic neutron scattering), to which end we have determined the energy and momentum dependence of the dipolar and quadrupolar dynamical structure factors.

{\it Acknowledgements ---}
We thank O. Janson, A. Tsirlin, H. Rosner, U. K. R\"o{\ss}ler, and M. Belesi for discussions and earlier collaborations on this compound. J.R. would like to thank K. Penc for useful discussions. We acknowledge support by the Deutsche Forschungsgemeinschaft (DFG) under the Emmy-Noether program.

\bibliographystyle{apsrev4-1}
\bibliography{Cu2OSeO3}

%Merlin.mbs v4.21 2009-07-09.
\begin{thebibliography}{10}%
\makeatletter
\providecommand \@ifxundefined [1]{%
 \ifx #1\undefined \expandafter \@firstoftwo
 \else \expandafter \@secondoftwo
\fi
}%
\providecommand \@ifnum [1]{%
 \ifnum #1\expandafter \@firstoftwo
 \else \expandafter \@secondoftwo
\fi
}%
\providecommand \enquote [1]{``#1''}%
\providecommand \bibnamefont  [1]{#1}%
\providecommand \bibfnamefont [1]{#1}%
\providecommand \citenamefont [1]{#1}%
\providecommand\href[0]{\@sanitize\@href}%
\providecommand\@href[1]{\endgroup\@@startlink{#1}\endgroup\@@href}%
\providecommand\@@href[1]{#1\@@endlink}%
\providecommand \@sanitize [0]{\begingroup\catcode`\&12\catcode`\#12\relax}%
\@ifxundefined \pdfoutput {\@firstoftwo}{%
 \@ifnum{\z@=\pdfoutput}{\@firstoftwo}{\@secondoftwo}%
}{%
 \providecommand\@@startlink[1]{\leavevmode\special{html:<a href="#1">}}%
 \providecommand\@@endlink[0]{\special{html:</a>}}%
}{%
 \providecommand\@@startlink[1]{%
  \leavevmode
  \pdfstartlink
   attr{/Border[0 0 1 ]/H/I/C[0 1 1]}%
   user{/Subtype/Link/A<</Type/Action/S/URI/URI(#1)>>}%
  \relax
 }%
 \providecommand\@@endlink[0]{\pdfendlink}%
}%
\providecommand \url  [0]{\begingroup\@sanitize \@url }%
\providecommand \@url [1]{\endgroup\@href {#1}{\urlprefix}}%
\providecommand \urlprefix [0]{URL }%
\providecommand \Eprint[0]{\href }%
\@ifxundefined \urlstyle {%
  \providecommand \doi [1]{doi:\discretionary{}{}{}#1}%
}{%
  \providecommand \doi [0]{doi:\discretionary{}{}{}\begingroup
  \urlstyle{rm}\Url }%
}%
\providecommand \doibase [0]{http://dx.doi.org/}%
\providecommand \Doi[1]{\href{\doibase#1}}%
\providecommand \bibAnnote [3]{%
  \BibitemShut{#1}%
  \begin{quotation}\noindent
    \textsc{Key:}\ #2\\\textsc{Annotation:}\ #3%
  \end{quotation}%
}%
\providecommand \bibAnnoteFile [2]{%
  \IfFileExists{#2}{\bibAnnote {#1} {#2} {\input{#2}}}{}%
}%
\providecommand \typeout [0]{\immediate \write \m@ne }%
\providecommand \selectlanguage [0]{\@gobble}%
\providecommand \bibinfo [0]{\@secondoftwo}%
\providecommand \bibfield [0]{\@secondoftwo}%
\providecommand \translation [1]{[#1]}%
\providecommand \BibitemOpen[0]{}%
\providecommand \bibitemStop [0]{}%
\providecommand \bibitemNoStop [0]{.\EOS\space}%
\providecommand \EOS [0]{\spacefactor3000\relax}%
\providecommand \BibitemShut [1]{\csname bibitem#1\endcsname}%
%</preamble>
\bibitem{Muhlbauer2009}%
  \BibitemOpen
  \bibfield{author}{%
  \bibinfo {author} {\bibfnamefont{S.}~\bibnamefont{M\"uhlbauer}}, \bibinfo
  {author} {\bibfnamefont{B.}~\bibnamefont{Binz}}, \bibinfo {author}
  {\bibfnamefont{F.}~\bibnamefont{Jonietz}}, \bibinfo {author}
  {\bibfnamefont{C.}~\bibnamefont{Pfleiderer}}, \bibinfo {author}
  {\bibfnamefont{A.}~\bibnamefont{Rosch}}, \bibinfo {author}
  {\bibfnamefont{A.}~\bibnamefont{Neubauer}}, \bibinfo {author}
  {\bibfnamefont{R.}~\bibnamefont{Georgii}},\ and\ \bibinfo {author}
  {\bibfnamefont{P.}~\bibnamefont{B\"oni}},\ }%
  \bibfield{journal}{%
  \Doi{10.1126/science.1166767}{\bibinfo {journal} {Science}}\ }%
  \textbf{\bibinfo {volume} {323}},\ \bibinfo {pages} {915} (\bibinfo {year}
  {2009})%
  \bibAnnoteFile{NoStop}{Muhlbauer2009}%
\bibitem{tonomura2012}%
  \BibitemOpen
  \bibfield{author}{%
  \bibinfo {author} {\bibfnamefont{A.}~\bibnamefont{Tonomura}}, \bibinfo
  {author} {\bibfnamefont{X.}~\bibnamefont{Yu}}, \bibinfo {author}
  {\bibfnamefont{K.}~\bibnamefont{Yanagisawa}}, \bibinfo {author}
  {\bibfnamefont{T.}~\bibnamefont{Matsuda}}, \bibinfo {author}
  {\bibfnamefont{Y.}~\bibnamefont{Onose}}, \bibinfo {author}
  {\bibfnamefont{N.}~\bibnamefont{Kanazawa}}, \bibinfo {author}
  {\bibfnamefont{H.~S.}\ \bibnamefont{Park}},\ and\ \bibinfo {author}
  {\bibfnamefont{Y.}~\bibnamefont{Tokura}},\ }%
  \bibfield{journal}{%
  \Doi{10.1021/nl300073m}{\bibinfo {journal} {Nano Letters}}\ }%
  \textbf{\bibinfo {volume} {12}},\ \bibinfo {pages} {1673} (\bibinfo {year}
  {2012})%
  \bibAnnoteFile{NoStop}{tonomura2012}%
\bibitem{Yu2010}%
  \BibitemOpen
  \bibfield{author}{%
  \bibinfo {author} {\bibfnamefont{X.~Z.}\ \bibnamefont{Yu}}, \bibinfo {author}
  {\bibfnamefont{Y.}~\bibnamefont{Onose}}, \bibinfo {author}
  {\bibfnamefont{N.}~\bibnamefont{Kanazawa}}, \bibinfo {author}
  {\bibfnamefont{J.~H.}\ \bibnamefont{Park}}, \bibinfo {author}
  {\bibfnamefont{J.~H.}\ \bibnamefont{Han}}, \bibinfo {author}
  {\bibfnamefont{Y.}~\bibnamefont{Matsui}}, \bibinfo {author}
  {\bibfnamefont{N.}~\bibnamefont{Nagaosa}},\ and\ \bibinfo {author}
  {\bibfnamefont{Y.}~\bibnamefont{Tokura}},\ }%
  \bibfield{journal}{%
  \Doi{10.1038/nature09124}{\bibinfo {journal} {Nature}}\ }%
  \textbf{\bibinfo {volume} {465}},\ \bibinfo {pages} {901} (\bibinfo {month}
  {Jun}\ \bibinfo {year} {2010}),\ \url{http://dx.doi.org/10.1038/nature09124}%
  \bibAnnoteFile{NoStop}{Yu2010}%
\bibitem{Muenzer2010}%
  \BibitemOpen
  \bibfield{author}{%
  \bibinfo {author} {\bibfnamefont{W.}~\bibnamefont{M\"unzer}}, \bibinfo
  {author} {\bibfnamefont{A.}~\bibnamefont{Neubauer}}, \bibinfo {author}
  {\bibfnamefont{T.}~\bibnamefont{Adams}}, \bibinfo {author}
  {\bibfnamefont{S.}~\bibnamefont{M\"uhlbauer}}, \bibinfo {author}
  {\bibfnamefont{C.}~\bibnamefont{Franz}}, \bibinfo {author}
  {\bibfnamefont{F.}~\bibnamefont{Jonietz}}, \bibinfo {author}
  {\bibfnamefont{R.}~\bibnamefont{Georgii}}, \bibinfo {author}
  {\bibfnamefont{P.}~\bibnamefont{B\"oni}}, \bibinfo {author}
  {\bibfnamefont{B.}~\bibnamefont{Pedersen}}, \bibinfo {author}
  {\bibfnamefont{M.}~\bibnamefont{Schmidt}}, \bibinfo {author}
  {\bibfnamefont{A.}~\bibnamefont{Rosch}},\ and\ \bibinfo {author}
  {\bibfnamefont{C.}~\bibnamefont{Pfleiderer}},\ }%
  \bibfield{journal}{%
  \Doi{10.1103/PhysRevB.81.041203}{\bibinfo {journal} {Phys. Rev. B}}\ }%
  \textbf{\bibinfo {volume} {81}},\ \bibinfo {pages} {041203} (\bibinfo {month}
  {Jan}\ \bibinfo {year} {2010}),\
  \url{http://link.aps.org/doi/10.1103/PhysRevB.81.041203}%
  \bibAnnoteFile{NoStop}{Muenzer2010}%
\bibitem{yu2012}%
  \BibitemOpen
  \bibfield{author}{%
  \bibinfo {author} {\bibfnamefont{X.}~\bibnamefont{Yu}}, \bibinfo {author}
  {\bibfnamefont{N.}~\bibnamefont{Kanazawa}}, \bibinfo {author}
  {\bibfnamefont{W.}~\bibnamefont{Zhang}}, \bibinfo {author}
  {\bibfnamefont{T.}~\bibnamefont{Nagai}}, \bibinfo {author}
  {\bibfnamefont{T.}~\bibnamefont{Hara}}, \bibinfo {author}
  {\bibfnamefont{K.}~\bibnamefont{Kimoto}}, \bibinfo {author}
  {\bibfnamefont{Y.}~\bibnamefont{Matsui}}, \bibinfo {author}
  {\bibfnamefont{Y.}~\bibnamefont{Onose}},\ and\ \bibinfo {author}
  {\bibfnamefont{Y.}~\bibnamefont{Tokura}},\ }%
  \bibfield{journal}{%
  \Doi{10.1038/ncomms1990}{\bibinfo {journal} {Nat Commun}}\ }%
  \textbf{\bibinfo {volume} {3}},\ \bibinfo {pages} {988} (\bibinfo {month}
  {Aug}\ \bibinfo {year} {2012}),\ \url{http://dx.doi.org/10.1038/ncomms1990}%
  \bibAnnoteFile{NoStop}{yu2012}%
\bibitem{Bogdanov1989a}%
  \BibitemOpen
  \bibfield{author}{%
  \bibinfo {author} {\bibfnamefont{A.~N.}\ \bibnamefont{Bogdanov}}\ and\
  \bibinfo {author} {\bibfnamefont{D.~A.}\ \bibnamefont{Yablonskii}},\ }%
  \bibfield{journal}{%
  \bibinfo {journal} {Sov. Phys. JETP}\ }%
  \textbf{\bibinfo {volume} {68}},\ \bibinfo {pages} {101} (\bibinfo {month}
  {Jan}\ \bibinfo {year} {1989}),\
  \url{http://jetp.ac.ru/cgi-bin/e/index/e/68/1/p101?a=list}%
  \bibAnnoteFile{NoStop}{Bogdanov1989a}%
\bibitem{Bogdanov1989b}%
  \BibitemOpen
  \bibfield{author}{%
  \bibinfo {author} {\bibfnamefont{A.~N.}\ \bibnamefont{Bogdanov}}\ and\
  \bibinfo {author} {\bibfnamefont{D.~A.}\ \bibnamefont{Yablonskii}},\ }%
  \bibfield{journal}{%
  \bibinfo {journal} {Sov. Phys. JETP}\ }%
  \textbf{\bibinfo {volume} {69}},\ \bibinfo {pages} {142} (\bibinfo {month}
  {July}\ \bibinfo {year} {1989}),\
  \url{http://www.jetp.ac.ru/cgi-bin/e/index/e/69/1/p142?a=list}%
  \bibAnnoteFile{NoStop}{Bogdanov1989b}%
\bibitem{Roessler2006}%
  \BibitemOpen
  \bibfield{author}{%
  \bibinfo {author} {\bibfnamefont{U.~K.}\ \bibnamefont{R\"o{\ss}ler}},
  \bibinfo {author} {\bibfnamefont{A.~N.}\ \bibnamefont{Bogdanov}},\ and\
  \bibinfo {author} {\bibfnamefont{C.}~\bibnamefont{Pfleiderer}},\ }%
  \bibfield{journal}{%
  \Doi{10.1038/nature05056}{\bibinfo {journal} {Nature}}\ }%
  \textbf{\bibinfo {volume} {442}},\ \bibinfo {pages} {797} (\bibinfo {month}
  {Aug}\ \bibinfo {year} {2006}),\ \url{http://dx.doi.org/10.1038/nature05056}%
  \bibAnnoteFile{NoStop}{Roessler2006}%
\bibitem{Abrikosov1957}%
  \BibitemOpen
  \bibfield{author}{%
  \bibinfo {author} {\bibfnamefont{A.~A.}\ \bibnamefont{Abrikosov}},\ }%
  \bibfield{journal}{%
  \bibinfo {journal} {Sov. Phys. JETP}\ }%
  \textbf{\bibinfo {volume} {5}},\ \bibinfo {pages} {1174} (\bibinfo {year}
  {1957})%
  \bibAnnoteFile{NoStop}{Abrikosov1957}%
\bibitem{Wright89}%
  \BibitemOpen
  \bibfield{author}{%
  \bibinfo {author} {\bibfnamefont{D.~C.}\ \bibnamefont{Wright}}\ and\ \bibinfo
  {author} {\bibfnamefont{N.~D.}\ \bibnamefont{Mermin}},\ }%
  \bibfield{journal}{%
  \Doi{10.1103/RevModPhys.61.385}{\bibinfo {journal} {Rev. Mod. Phys.}}\ }%
  \textbf{\bibinfo {volume} {61}},\ \bibinfo {pages} {385} (\bibinfo {month}
  {Apr}\ \bibinfo {year} {1989})%
  \bibAnnoteFile{NoStop}{Wright89}%
\bibitem{Seki2012Sc}%
  \BibitemOpen
  \bibfield{author}{%
  \bibinfo {author} {\bibfnamefont{S.}~\bibnamefont{Seki}}, \bibinfo {author}
  {\bibfnamefont{X.~Z.}\ \bibnamefont{Yu}}, \bibinfo {author}
  {\bibfnamefont{S.}~\bibnamefont{Ishiwata}},\ and\ \bibinfo {author}
  {\bibfnamefont{Y.}~\bibnamefont{Tokura}},\ }%
  \bibfield{journal}{%
  \Doi{10.1126/science.1214143}{\bibinfo {journal} {Science}}\ }%
  \textbf{\bibinfo {volume} {336}},\ \bibinfo {pages} {198} (\bibinfo {year}
  {2012})%
  \bibAnnoteFile{NoStop}{Seki2012Sc}%
\bibitem{Bos2008}%
  \BibitemOpen
  \bibfield{author}{%
  \bibinfo {author} {\bibfnamefont{J.-W.~G.}\ \bibnamefont{Bos}}, \bibinfo
  {author} {\bibfnamefont{C.~V.}\ \bibnamefont{Colin}},\ and\ \bibinfo {author}
  {\bibfnamefont{T.~T.~M.}\ \bibnamefont{Palstra}},\ }%
  \bibfield{journal}{%
  \Doi{10.1103/PhysRevB.78.094416}{\bibinfo {journal} {Phys. Rev. B}}\ }%
  \textbf{\bibinfo {volume} {78}},\ \bibinfo {pages} {094416} (\bibinfo {month}
  {Sep}\ \bibinfo {year} {2008}),\
  \url{http://link.aps.org/doi/10.1103/PhysRevB.78.094416}%
  \bibAnnoteFile{NoStop}{Bos2008}%
\bibitem{Belesi2012}%
  \BibitemOpen
  \bibfield{author}{%
  \bibinfo {author} {\bibfnamefont{M.}~\bibnamefont{Belesi}}, \bibinfo {author}
  {\bibfnamefont{I.}~\bibnamefont{Rousochatzakis}}, \bibinfo {author}
  {\bibfnamefont{M.}~\bibnamefont{Abid}}, \bibinfo {author}
  {\bibfnamefont{U.~K.}\ \bibnamefont{R\"o\ss{}ler}}, \bibinfo {author}
  {\bibfnamefont{H.}~\bibnamefont{Berger}},\ and\ \bibinfo {author}
  {\bibfnamefont{J.-P.}\ \bibnamefont{Ansermet}},\ }%
  \bibfield{journal}{%
  \Doi{10.1103/PhysRevB.85.224413}{\bibinfo {journal} {Phys. Rev. B}}\ }%
  \textbf{\bibinfo {volume} {85}},\ \bibinfo {pages} {224413} (\bibinfo {month}
  {Jun}\ \bibinfo {year} {2012}),\
  \url{http://link.aps.org/doi/10.1103/PhysRevB.85.224413}%
  \bibAnnoteFile{NoStop}{Belesi2012}%
\bibitem{SekiarXiv}%
  \BibitemOpen
  \bibfield{author}{%
  \bibinfo {author} {\bibfnamefont{S.}~\bibnamefont{Seki}}, \bibinfo {author}
  {\bibfnamefont{S.}~\bibnamefont{Ishiwata}},\ and\ \bibinfo {author}
  {\bibfnamefont{Y.}~\bibnamefont{Tokura}},\ }%
  \bibfield{journal}{%
  \Doi{10.1103/PhysRevB.86.060403}{\bibinfo {journal} {Phys. Rev. B}}\ }%
  \textbf{\bibinfo {volume} {86}},\ \bibinfo {pages} {060403} (\bibinfo {month}
  {Aug}\ \bibinfo {year} {2012}),\
  \url{http://link.aps.org/doi/10.1103/PhysRevB.86.060403}%
  \bibAnnoteFile{NoStop}{SekiarXiv}%
\bibitem{White2012}%
  \BibitemOpen
  \bibfield{author}{%
  \bibinfo {author} {\bibfnamefont{J.~S.}\ \bibnamefont{White}}, \bibinfo
  {author} {\bibfnamefont{I.}~\bibnamefont{Levatic}}, \bibinfo {author}
  {\bibfnamefont{A.~A.}\ \bibnamefont{Omrani}}, \bibinfo {author}
  {\bibfnamefont{N.}~\bibnamefont{Egetenmeyer}}, \bibinfo {author}
  {\bibfnamefont{K.}~\bibnamefont{Prsa}}, \bibinfo {author}
  {\bibfnamefont{I.}~\bibnamefont{Zivkovic}}, \bibinfo {author}
  {\bibfnamefont{J.~L.}\ \bibnamefont{Gavilano}}, \bibinfo {author}
  {\bibfnamefont{J.}~\bibnamefont{Kohlbrecher}}, \bibinfo {author}
  {\bibfnamefont{M.}~\bibnamefont{Bartkowiak}}, \bibinfo {author}
  {\bibfnamefont{H.}~\bibnamefont{Berger}},\ and\ \bibinfo {author}
  {\bibfnamefont{H.~M.}\ \bibnamefont{Ronnow}},\ }%
  \bibfield{journal}{%
  \Doi{10.1088/0953-8984/24/43/432201}{\bibinfo {journal} {J. Phys.: Condens.
  Matter}}\ }%
  \textbf{\bibinfo {volume} {24}},\ \bibinfo {pages} {432201} (\bibinfo {year}
  {2012})%
  \bibAnnoteFile{NoStop}{White2012}%
\bibitem{Lin2013}%
  \BibitemOpen
  \bibfield{author}{%
  \bibinfo {author} {\bibfnamefont{S.-Z.}\ \bibnamefont{Lin}}, \bibinfo
  {author} {\bibfnamefont{C.}~\bibnamefont{Reichhardt}}, \bibinfo {author}
  {\bibfnamefont{C.~D.}\ \bibnamefont{Batista}},\ and\ \bibinfo {author}
  {\bibfnamefont{A.}~\bibnamefont{Saxena}}}%
   (\bibinfo {year} {2013}),\
  \Eprint{http://arxiv.org/abs/arXiv:1301.5963}{arXiv:1301.5963}%
  \bibAnnoteFile{NoStop}{Lin2013}%
\bibitem{Fert2013}%
  \BibitemOpen
  \bibfield{author}{%
  \bibinfo {author} {\bibfnamefont{A.}~\bibnamefont{Fert}}, \bibinfo {author}
  {\bibfnamefont{V.}~\bibnamefont{Cros}},\ and\ \bibinfo {author}
  {\bibfnamefont{J.}~\bibnamefont{Sampaio}},\ }%
  \bibfield{journal}{%
  \Doi{10.1038/nnano.2013.29}{\bibinfo {journal} {Nature Nanotech.}}\ }%
  \textbf{\bibinfo {volume} {8}},\ \bibinfo {pages} {152} (\bibinfo {year}
  {2013})%
  \bibAnnoteFile{NoStop}{Fert2013}%
\bibitem{Yang2012}%
  \BibitemOpen
  \bibfield{author}{%
  \bibinfo {author} {\bibfnamefont{J.~H.}\ \bibnamefont{Yang}}, \bibinfo
  {author} {\bibfnamefont{Z.~L.}\ \bibnamefont{Li}}, \bibinfo {author}
  {\bibfnamefont{X.~Z.}\ \bibnamefont{Lu}}, \bibinfo {author}
  {\bibfnamefont{M.-H.}\ \bibnamefont{Whangbo}}, \bibinfo {author}
  {\bibfnamefont{S.-H.}\ \bibnamefont{Wei}}, \bibinfo {author}
  {\bibfnamefont{X.~G.}\ \bibnamefont{Gong}},\ and\ \bibinfo {author}
  {\bibfnamefont{H.~J.}\ \bibnamefont{Xiang}},\ }%
  \bibfield{journal}{%
  \Doi{10.1103/PhysRevLett.109.107203}{\bibinfo {journal} {Phys. Rev. Lett.}}\
  }%
  \textbf{\bibinfo {volume} {109}},\ \bibinfo {pages} {107203} (\bibinfo
  {month} {Sep}\ \bibinfo {year} {2012}),\
  \url{http://link.aps.org/doi/10.1103/PhysRevLett.109.107203}%
  \bibAnnoteFile{NoStop}{Yang2012}%
\bibitem{Janson2013}%
  \BibitemOpen
  \bibfield{author}{%
  \bibinfo {author} {\bibfnamefont{O.}~\bibnamefont{Janson}}, \bibinfo {author}
  {\bibfnamefont{I.}~\bibnamefont{Rousochatzakis}}, \bibinfo {author}
  {\bibfnamefont{A.~A.}\ \bibnamefont{Tsirlin}}, \bibinfo {author}
  {\bibfnamefont{M.}~\bibnamefont{Belesi}}, \bibinfo {author}
  {\bibfnamefont{A.~A.}\ \bibnamefont{Leonov}}, \bibinfo {author}
  {\bibfnamefont{U.~K.}\ \bibnamefont{R\"{o}{\ss}ler}}, \bibinfo {author}
  {\bibfnamefont{J.}~\bibnamefont{van~den Brink}},\ and\ \bibinfo {author}
  {\bibfnamefont{H.}~\bibnamefont{Rosner}},\ }%
  \bibfield{journal}{%
  \Doi{10.1038/ncomms6376}{\bibinfo {journal} {Nat Commun}}\ }%
  \textbf{\bibinfo {volume} {5}},\ \bibinfo {pages} {6376} (\bibinfo {year}
  {2014})%
  \bibAnnoteFile{NoStop}{Janson2013}%
\bibitem{ozerov2014}%
  \BibitemOpen
  \bibfield{author}{%
  \bibinfo {author} {\bibfnamefont{M.}~\bibnamefont{Ozerov}}, \bibinfo {author}
  {\bibfnamefont{J.}~\bibnamefont{Romh\'anyi}}, \bibinfo {author}
  {\bibfnamefont{M.}~\bibnamefont{Belesi}}, \bibinfo {author}
  {\bibfnamefont{H.}~\bibnamefont{Berger}}, \bibinfo {author}
  {\bibfnamefont{J.-P.}\ \bibnamefont{Ansermet}}, \bibinfo {author}
  {\bibfnamefont{J.}~\bibnamefont{van~den Brink}}, \bibinfo {author}
  {\bibfnamefont{J.}~\bibnamefont{Wosnitza}}, \bibinfo {author}
  {\bibfnamefont{S.~A.}\ \bibnamefont{Zvyagin}},\ and\ \bibinfo {author}
  {\bibfnamefont{I.}~\bibnamefont{Rousochatzakis}},\ }%
  \bibfield{journal}{%
  \Doi{10.1103/PhysRevLett.113.157205}{\bibinfo {journal} {Phys. Rev. Lett.}}\
  }%
  \textbf{\bibinfo {volume} {113}},\ \bibinfo {pages} {157205} (\bibinfo
  {month} {Oct}\ \bibinfo {year} {2014})%
  \bibAnnoteFile{NoStop}{ozerov2014}%
\bibitem{Gnezdilov2010}%
  \BibitemOpen
  \bibfield{author}{%
  \bibinfo {author} {\bibfnamefont{V.~P.}\ \bibnamefont{Gnezdilov}}, \bibinfo
  {author} {\bibfnamefont{K.~V.}\ \bibnamefont{Lamonova}}, \bibinfo {author}
  {\bibfnamefont{Y.~G.}\ \bibnamefont{Pashkevich}}, \bibinfo {author}
  {\bibfnamefont{P.}~\bibnamefont{Lemmens}}, \bibinfo {author}
  {\bibfnamefont{H.}~\bibnamefont{Berger}}, \bibinfo {author}
  {\bibfnamefont{F.}~\bibnamefont{Bussy}},\ and\ \bibinfo {author}
  {\bibfnamefont{S.~L.}\ \bibnamefont{Gnatchenko}},\ }%
  \bibfield{journal}{%
  \Doi{http://dx.doi.org/10.1063/1.3455808}{\bibinfo {journal} {Low Temperature
  Physics}}\ }%
  \textbf{\bibinfo {volume} {36}},\ \bibinfo {pages} {550} (\bibinfo {year}
  {2010}),\ \url{http://link.aip.org/link/?LTP/36/550/1}%
  \bibAnnoteFile{NoStop}{Gnezdilov2010}%
\bibitem{Miller2010}%
  \BibitemOpen
  \bibfield{author}{%
  \bibinfo {author} {\bibfnamefont{K.~H.}\ \bibnamefont{Miller}}, \bibinfo
  {author} {\bibfnamefont{X.~S.}\ \bibnamefont{Xu.}}, \bibinfo {author}
  {\bibfnamefont{H.}~\bibnamefont{Berger}}, \bibinfo {author}
  {\bibfnamefont{E.~S.}\ \bibnamefont{Knowles}}, \bibinfo {author}
  {\bibfnamefont{D.~J.}\ \bibnamefont{Arenas}}, \bibinfo {author}
  {\bibfnamefont{M.~W.}\ \bibnamefont{Meisel}},\ and\ \bibinfo {author}
  {\bibfnamefont{D.~B.}\ \bibnamefont{Tanner}},\ }%
  \bibfield{journal}{%
  \Doi{10.1103/PhysRevB.82.144107}{\bibinfo {journal} {Phys. Rev. B}}\ }%
  \textbf{\bibinfo {volume} {82}},\ \bibinfo {pages} {144107} (\bibinfo {month}
  {Oct}\ \bibinfo {year} {2010}),\
  \url{http://link.aps.org/doi/10.1103/PhysRevB.82.144107}%
  \bibAnnoteFile{NoStop}{Miller2010}%
\bibitem{suppl}%
  \BibitemOpen
  \bibinfo {note} {See supplemental material for details on the eigenspectra of
  $\mathcal{H}^{(t)}_0$ and $\mathcal{H}^{(t)}_{\sf TMF}$ and the multiboson
  expansion method.}%
  \bibAnnoteFile{Stop}{suppl}%
\bibitem{Belesi2010}%
  \BibitemOpen
  \bibfield{author}{%
  \bibinfo {author} {\bibfnamefont{M.}~\bibnamefont{Belesi}}, \bibinfo {author}
  {\bibfnamefont{I.}~\bibnamefont{Rousochatzakis}}, \bibinfo {author}
  {\bibfnamefont{H.~C.}\ \bibnamefont{Wu}}, \bibinfo {author}
  {\bibfnamefont{H.}~\bibnamefont{Berger}}, \bibinfo {author}
  {\bibfnamefont{I.~V.}\ \bibnamefont{Shvets}}, \bibinfo {author}
  {\bibfnamefont{F.}~\bibnamefont{Mila}},\ and\ \bibinfo {author}
  {\bibfnamefont{J.~P.}\ \bibnamefont{Ansermet}},\ }%
  \bibfield{journal}{%
  \Doi{10.1103/PhysRevB.82.094422}{\bibinfo {journal} {Phys. Rev. B}}\ }%
  \textbf{\bibinfo {volume} {82}},\ \bibinfo {pages} {094422} (\bibinfo {month}
  {Sep}\ \bibinfo {year} {2010}),\
  \url{http://link.aps.org/doi/10.1103/PhysRevB.82.094422}%
  \bibAnnoteFile{NoStop}{Belesi2010}%
\bibitem{Papanicolaou1984}%
  \BibitemOpen
  \bibfield{author}{%
  \bibinfo {author} {\bibfnamefont{N.}~\bibnamefont{Papanicolaou}},\ }%
  \bibfield{journal}{%
  \Doi{10.1016/0550-3213(84)90268-2}{\bibinfo {journal} {Nuclear Physics B}},\
  \bibinfo {pages} {281 }\ }%
  ISSN \bibinfo {issn} {0550-3213}%
  \bibAnnoteFile{NoStop}{Papanicolaou1984}%
\bibitem{Papanicolaou1988}%
  \BibitemOpen
  \bibfield{author}{%
  \bibinfo {author} {\bibfnamefont{N.}~\bibnamefont{Papanicolaou}},\ }%
  \bibfield{journal}{%
  \Doi{10.1016/0550-3213(88)90073-9}{\bibinfo {journal} {Nuclear Physics B}},\
  \bibinfo {pages} {367 }\ }%
  ISSN \bibinfo {issn} {0550-3213}%
  \bibAnnoteFile{NoStop}{Papanicolaou1988}%
\bibitem{Onufrieva1985}%
  \BibitemOpen
  \bibfield{author}{%
  \bibinfo {author} {\bibfnamefont{F.}~\bibnamefont{Onufrieva}},\ }%
  \bibfield{journal}{%
  \bibinfo {journal} {Zh. Eksp. Teor. Fiz.}\ }%
  \textbf{\bibinfo {volume} {89}},\ \bibinfo {pages} {2270} (\bibinfo {month}
  {DEC}\ \bibinfo {year} {1985}),\ ISSN \bibinfo {issn} {0044-4510}%
  \bibAnnoteFile{NoStop}{Onufrieva1985}%
\bibitem{Chubukov1990}%
  \BibitemOpen
  \bibfield{author}{%
  \bibinfo {author} {\bibfnamefont{A.~V.}\ \bibnamefont{Chubukov}},\ }%
  \bibinfo {journal} {J. Phys. Condens. Matter},\ \bibinfo {pages} {1593}%
  \bibAnnoteFile{NoStop}{Chubukov1990}%
\bibitem{Holstein1940}%
  \BibitemOpen
\bibfield{journal}{%
    }%
  \bibfield{author}{%
  \bibinfo {author} {\bibfnamefont{T.}~\bibnamefont{Holstein}}\ and\ \bibinfo
  {author} {\bibfnamefont{H.}~\bibnamefont{Primakoff}},\ }%
  \bibfield{journal}{%
  \Doi{10.1103/PhysRev.58.1098}{\bibinfo {journal} {Phys. Rev.}}\ }%
  \textbf{\bibinfo {volume} {58}},\ \bibinfo {pages} {1098} (\bibinfo {month}
  {Dec}\ \bibinfo {year} {1940}),\
  \url{http://link.aps.org/doi/10.1103/PhysRev.58.1098}%
  \bibAnnoteFile{NoStop}{Holstein1940}%
\bibitem{Anderson1952}%
  \BibitemOpen
  \bibfield{author}{%
  \bibinfo {author} {\bibfnamefont{P.~W.}\ \bibnamefont{Anderson}},\ }%
  \bibfield{journal}{%
  \Doi{10.1103/PhysRev.86.694}{\bibinfo {journal} {Phys. Rev.}}\ }%
  \textbf{\bibinfo {volume} {86}},\ \bibinfo {pages} {694} (\bibinfo {month}
  {Jun}\ \bibinfo {year} {1952}),\
  \url{http://link.aps.org/doi/10.1103/PhysRev.86.694}%
  \bibAnnoteFile{NoStop}{Anderson1952}%
\bibitem{Kubo1952}%
  \BibitemOpen
  \bibfield{author}{%
  \bibinfo {author} {\bibfnamefont{R.}~\bibnamefont{Kubo}},\ }%
  \bibfield{journal}{%
  \Doi{10.1103/PhysRev.87.568}{\bibinfo {journal} {Phys. Rev.}}\ }%
  \textbf{\bibinfo {volume} {87}},\ \bibinfo {pages} {568} (\bibinfo {month}
  {Aug}\ \bibinfo {year} {1952}),\
  \url{http://link.aps.org/doi/10.1103/PhysRev.87.568}%
  \bibAnnoteFile{NoStop}{Kubo1952}%
\bibitem{Romhanyi2011}%
  \BibitemOpen
  \bibfield{author}{%
  \bibinfo {author} {\bibfnamefont{J.}~\bibnamefont{Romh\'anyi}}, \bibinfo
  {author} {\bibfnamefont{K.}~\bibnamefont{Totsuka}},\ and\ \bibinfo {author}
  {\bibfnamefont{K.}~\bibnamefont{Penc}},\ }%
  \bibfield{journal}{%
  \Doi{10.1103/PhysRevB.83.024413}{\bibinfo {journal} {Phys. Rev. B}}\ }%
  \textbf{\bibinfo {volume} {83}},\ \bibinfo {pages} {024413} (\bibinfo {month}
  {Jan}\ \bibinfo {year} {2011}),\
  \url{http://link.aps.org/doi/10.1103/PhysRevB.83.024413}%
  \bibAnnoteFile{NoStop}{Romhanyi2011}%
\bibitem{Carretta2010}%
  \BibitemOpen
  \bibfield{author}{%
  \bibinfo {author} {\bibfnamefont{S.}~\bibnamefont{Carretta}}, \bibinfo
  {author} {\bibfnamefont{P.}~\bibnamefont{Santini}}, \bibinfo {author}
  {\bibfnamefont{R.}~\bibnamefont{Caciuffo}},\ and\ \bibinfo {author}
  {\bibfnamefont{G.}~\bibnamefont{Amoretti}},\ }%
  \bibfield{journal}{%
  \Doi{10.1103/PhysRevLett.105.167201}{\bibinfo {journal} {Phys. Rev. Lett.}}\
  }%
  \textbf{\bibinfo {volume} {105}},\ \bibinfo {pages} {167201} (\bibinfo
  {month} {Oct}\ \bibinfo {year} {2010}),\
  \url{http://link.aps.org/doi/10.1103/PhysRevLett.105.167201}%
  \bibAnnoteFile{NoStop}{Carretta2010}%
\bibitem{Shiina2003}%
  \BibitemOpen
  \bibfield{author}{%
  \bibinfo {author} {\bibfnamefont{R.}~\bibnamefont{Shiina}}, \bibinfo {author}
  {\bibfnamefont{H.}~\bibnamefont{Shiba}}, \bibinfo {author}
  {\bibfnamefont{P.}~\bibnamefont{Thalmeier}}, \bibinfo {author}
  {\bibfnamefont{A.}~\bibnamefont{Takahashi}},\ and\ \bibinfo {author}
  {\bibfnamefont{O.}~\bibnamefont{Sakai}},\ }%
  \Doi{10.1143/JPSJ.72.1216}{\bibinfo {journal} {J. Phys. Soc. Jpn.}},\
  \bibinfo {pages} {1216}%
  \bibAnnoteFile{NoStop}{Shiina2003}%
\bibitem{nematicreview2011}%
  \BibitemOpen
\bibfield{journal}{%
    }%
  \bibfield{author}{%
  \bibinfo {author} {\bibfnamefont{K.}~\bibnamefont{Penc}}\ and\ \bibinfo
  {author} {\bibfnamefont{A.~M.}\ \bibnamefont{L\"auchli}},\ }%
  in\ \emph{\bibinfo {booktitle} {Introduction to Frustrated Magnetism}},\
  \bibinfo {series and number} {Springer Series in Solid-State Sciences},\
  \bibinfo {editor} {edited by\ \bibinfo {editor}
  {\bibfnamefont{C.}~\bibnamefont{Lacroix}}, \bibinfo {editor}
  {\bibfnamefont{P.}~\bibnamefont{Mendels}},\ and\ \bibinfo {editor}
  {\bibfnamefont{F.}~\bibnamefont{Mila}}}\ (\bibinfo {publisher} {Springer
  Berlin Heidelberg})\ pp.\ \bibinfo {pages} {331--362},\ ISBN \bibinfo {isbn}
  {978-3-642-10589-0}%
  \bibAnnoteFile{NoStop}{nematicreview2011}%
\bibitem{Blume1969}%
  \BibitemOpen
  \bibfield{author}{%
  \bibinfo {author} {\bibfnamefont{M.}~\bibnamefont{Blume}}\ and\ \bibinfo
  {author} {\bibfnamefont{Y.~Y.}\ \bibnamefont{Hsieh}},\ }%
  \bibinfo {journal} {J. Appl. Phys.},\ \bibinfo {pages} {1249}%
  \bibAnnoteFile{NoStop}{Blume1969}%
\bibitem{modeA}%
  \BibitemOpen
\bibfield{journal}{%
    }%
  \bibinfo {note} {While the debate on the origin of the Raman line at
  84.6\,cm$^{-1}$ remains open~\cite{Gnezdilov2010,Miller2010}, the
  extrapolation of the ESR data for mode A gives an energy of
  95\,cm$^{-1}$~\cite{ozerov2014}, almost on top of our theoretical
  calculations for mode A.}%
  \bibAnnoteFile{Stop}{modeA}%
\end{thebibliography}%

\clearpage

\end{document}